\documentstyle[prl,aps,multicol,epsf]{revtex}
\def\inseps#1#2{\def\epsfsize##1##2{#2##1} \centerline{\epsfbox{#1}}}

\begin{document}
\draft
\title{Adsorption-like Collapse of Diblock Copolymers}
\author{Enzo Orlandini $^1$, Flavio Seno $^1$, and Attilio L. Stella $^{1,2}$}
\address{$^1$ INFM-Dipartimento di Fisica,\\
Universit\`a di Padova, I-35131 Padova, Italy\\
$^2$ The Abdus Salam ICTP, P. O. Box 563, I-34100, Trieste, Italy,\\
Sezione INFN, Universit\`a di Padova, I-35131 Padova, Italy }
\maketitle

\begin{abstract}
A linear copolymer made of two reciprocally attracting $N$-monomer blocks
collapses to a compact phase through a novel transition,
whose exponents are determined with extensive MC simulations in 
two and three dimensions. In the former case, an identification with
the statistical geometry of suitable percolation paths allows to 
predict that the number of contacts between the blocks grows like $N^{9/16}$. 
In the compact phase the blocks are mixed and, in two dimensions,
also zipped, in such a way to form a spiral, double chain structure.
\end{abstract}

\pacs{05.70.Jk,64.60.Ak,64.60.Kw,36.20.-r}

%\vspace{1cm}

\begin{multicols}{2}
\narrowtext

Our knowledge of the conformational properties in solution of single, isolated
polymers with inhomogeneous backbone sequence, 
is still extremely limited\cite{GOP98}. 
A major challenge in statistical physics consists
in describing the possible 
conformational transitions of such
systems, their universal scaling behaviors, and the nature of the different
phases they connect, in
finite dimensional situations. In the last decades a similar program has been 
accomplished, to a substantial extent, for homopolymers\cite{Carlo98},
while for heteropolymers, so far, most insight is limited
to approaches of mean field type\cite{GOP98,SS97}.

Block copolymers\cite{copolymer,SS97} are interesting as relatively elementary 
members of the large family of heteropolymers, like proteins or polyampholytes,
characterized by the above inhomogeneity.
These copolymers are important as interface stabilizing 
agents, and can display intriguing phenomena of microphase 
separation, a major topic in soft condensed matter research.

The status of the art, as far as studies of hetero- and co-polymer 
conformational transitions are concerned, suggests us to investigate in detail,
without resort to mean field approximations,
prototype problems involving relatively simple molecular architectures.
Of particular interest will be those transitions which
possibly reveal peculiar to systems with chain inhomogeneity, 
without counterpart in the homopolymer case.

In this Letter we address one of such problems, the collapse
from high temperature ($T$) swollen, to low $T$ compact state
of a copolymer whose two equally long blocks, A and B, attract each other
with short-range forces. A physical realization could be
that of oppositely charged A and B, immersed in a screening solvent.
Another case, involving AA and BB attraction, rather than
repulsion, is one in which the monomers of A are able to
establish hydrogen bonds with those of B, which add to the
Van der Waals forces, thus creating extra AB attractive interactions.
In first instance we consider here the effect of AB attractive 
interactions acting alone.
In both two dimensions (2D) and 3D, the collapsed phase has
an approximate alternating, periodic structure,
as far as the space distribution of A and B
monomers is concerned. 
We determine numerically, and predict exactly in 2D,
the exponents of this collapse,
showing that it has qualitative features in common with adsorption
phenomena\cite{DBL93}. Nevertheless, the universality class coincides
with neither that of adsorption\cite{DBL93}, 
nor that of theta collapse of homopolymers\cite{DS87}.

We model the diblock copolymer by a self-avoiding walk (SAW) $w$ 
of length $|w|=2N$ steps (monomers) on square and cubic lattice. 
The walk $w$ consists of $N$ consecutive
monomers of type A ($w_A$), followed by $N$ monomers of type B ($w_B$).
The Hamiltonian takes the form
\begin{equation}
H(w )=-\sum_{i\in w_A, j\in w_B} \epsilon
\end{equation}
where $\epsilon$ is a positive energy, and the sum is supposed to
run over pairs $<i,j>$ of lattice sites visited by the copolymer
( the AB junction excluded), which are nearest neighbors 
(n.n.). We will also consider the
possible inclusion in $H$ of extra terms representing analogous n.n. 
AA and BB repulsive or attractive potentials\cite{OSS99}.

The scaling regime of a polymer can be conveniently described
by the $\nu$ exponent governing the asymptotic behavior of the
canonical average radius of gyration, 
\begin{equation}
\langle R_g \rangle=\frac{\sum_{w} \exp(-H(w)/T) R(w)}{\sum_{w}\exp(-H(w)/T)} \sim N^{\nu},
\end{equation}
where the sums extend to all $2N$-step configurations $w$ of the
copolymer, with radius $R(w)$ relative to the center of mass.
In the swollen, high $T$ regime one should expect $\nu = \nu_{SAW}$, with 
$\nu_{SAW} = 3/4$\cite{N82}
 and $0.588$\cite{GZ98} in 2D and 3D, respectively. In 
a compact, low $T$ phase, $\nu = 1/d$.
Another important physical quantity is the specific heat 
$C=\frac{1}{N}\partial<H>/\partial T$,
which at a conformational transition between high $T$ and low $T$ regimes
is expected to obey a scaling of the form:
\begin{equation}
C \sim N^{2\phi-1} F((T-T_c)N^{\phi})
\end{equation}
for large $N$ and for $T$ close to the transition temperature $T_c$, 
$F$ being a suitable scaling function.
For $T=T_c$, $\nu$ should take a peculiar value, $\nu_c$.

By extensive Monte Carlo sampling we computed $\langle R_g\rangle$,
$C$ and $\langle H \rangle$ of the copolymer for $N$ up to $800$ in both
2D and 3D. 
The sampling was based on a multiple Markov chain (MMC) method, which
has been shown to be particularly suitable to deal with polymers at
low $T$\cite{TJOW96}. Several
Markov chains, each one designed to sample at a different $T$, 
are generated in parallel, by using both pivot\cite{MS87} and
local moves\cite{VS61}. The sampling
at low $T$ is then considerably
enriched by swapping configurations between
chains contiguous in $T$. Since each Markov chain is
ergodic, so is the composite one\cite{TJOW96}.
The swapping procedure dramatically reduces the correlations
within each chain, and produces little CPU waste time since,
in any case, one is interested in obtaining data at several 
$T$'s\cite{TJOW96}. Both in 2D and 3D the simulations have been performed
by using 25 parallel chains covering a range from $T = \infty$
to deeply inside the collapsed phase ($T=0.5$).

\begin{figure}
\vskip -0.25 truecm
\inseps{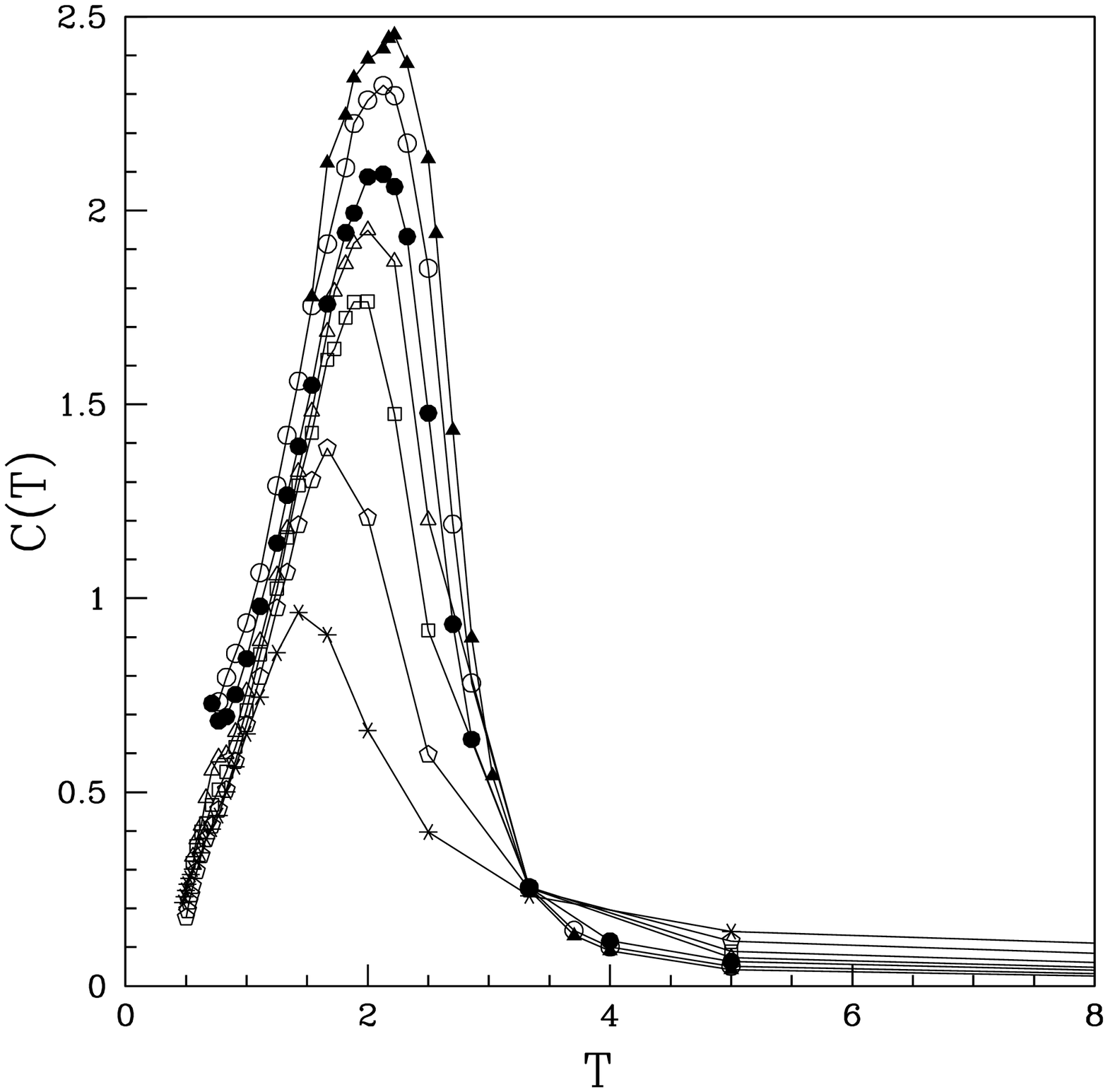}{0.25}
\vskip 0.1 truecm
\caption{Specific heat as a function of T for  $N=50$ ($*$),
 $100$ ($\diamond$), $200$ ($\scriptstyle \Box$),$300$ ($\scriptstyle \Delta$),
$400$ ($ \bullet$), $600$ ($\circ$) and $800$ (filled triangles). }
\label{fig1}
\end{figure}

Our data for $C$ ( Fig.\ref{fig1}) signal clearly
the presence of a transition, with peaks in the $T$-dependence
sharpening and growing
with $N$, consistently with Eq.(3).
The corresponding $T_c$ and the crossover exponent $\phi$
could be deduced from the $N$ dependence of the height, $h(N)$, and position,
$T_c(N)$, of these peaks. Indeed, in the scaling limit we expect
$h(N)\sim N^{2\phi-1}$ and $T_c(N)-T_c(\infty)\sim N^{-\phi}$,
for increasing $N$ ($T_c(\infty)=T_c$).

\begin{figure}
\vskip -0.5 truecm
\inseps{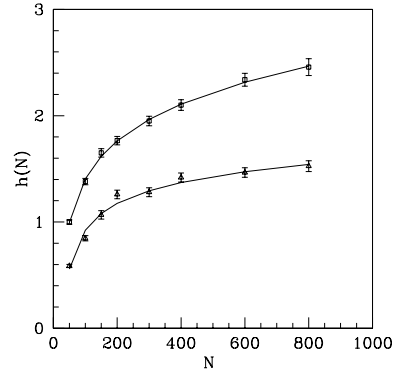}{0.25}
%\vskip 0.1 truecm
\caption{Fits of $h(N)$ in 2D (triangles) and 3D (squares).}
\label{fig2}
\end{figure}

\begin{figure}
\vskip -0.4truecm
\inseps{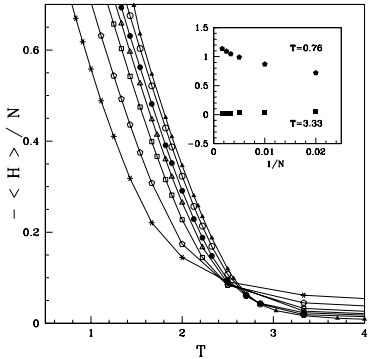}{0.25}
\caption{$\langle H \rangle / N$ vs T in 3D for $N=50$ ($*$),
 $100$ ($\diamond$), $200$ ($\scriptstyle \Box$),$300$ ($\scriptstyle \Delta$),
$400$ ($ \bullet$), $600$ ($\circ$) and $800$ (filled triangles). 
The inset shows the convergence
of $\langle H \rangle / N$ to $0$ for $T>T_c$ and to a constant for $T<T_c$. }
\label{fig3}
\end{figure}

Since a linear least squares fit of the log of the
asymptotic form of $h$ gives a very large $\chi^2$ statistical error, 
we consider $h(N) = A N^{2\phi-1}(1+B/N)$
where a scaling correction $1/N$ is included. The least squares
fit in this case gives a lower $\chi^2$ 
and we obtained (see Fig.\ref{fig2}) $\phi = 0.56 \pm 0.02$ in 2D 
and $\phi = 0.60 \pm 0.01$ in 3D.
These values of $\phi$ allowed to extrapolate
$T_c(\infty)/\epsilon= 1.5\pm 0.2$ and $T_c(\infty)/\epsilon= 2.9 \pm 0.2$, in 2D and 3D,
respectively. From log-log plots of
$\langle R_g\rangle$ vs. $N$ we also estimated 
the $\nu$ exponents at several temperatures. 
At $T=T_c$, as determined above, we obtain
$\nu_c = 0.748 \pm 0.008 $ in 2D, and $\nu_c = 0.583 \pm 0.007$ in 3D. 
In fact $\nu$ stays approximately constant for $T\geq T_c$.
Only for
$T<T_c$, $\nu$ decreases, approaching values consistent with a 
compact phase ($\nu \sim 1/d$) for low $T$.
The $\nu_c$ estimates appear fully compatible
with the exponents appropiate to a swollen SAW. 
In this respect the new collapse is analogous to a polymer adsorption on an 
attractive, impenetrable wall\cite{DBL93}. 
Indeed, at an adsorption transition one 
also finds $\nu_{ads}=\nu_{SAW}$.
However, the crossover exponents $\phi$
determined here are definitely different from 
those describing the growth with $N$ of the 
number of polymer-wall contacts in
adsorption ($\phi_{ads}=1/2$ and 
$\phi_{ads}= 0.496\pm 0.004$
in 2D\cite{DBL93} and 3D\cite{HG94}, respectively).
Further analogies with adsorption are revealed by
the large $N$ behavior of the average energy (or $\langle N_{AB}\rangle$)
per monomer. Fig.\ref{fig3} indicates
crossings of the various $ \langle H \rangle /N$ curves.
The temperature range at which the various crossings concentrate
is a signal of the transition and is consistent with the above estimate
of $T_c$.
Moreover
$ \langle H \rangle /N \to 0$ for $T>T_c$ and
$\langle H \rangle /N \to \hbox{const}$ for $T<T_c$
(inset of Fig. \ref{fig3}).
This behavior is completely different from what is found
for the theta collapse, where the energy density curves
do not cross each other, and $\langle H \rangle /N$ is asymptotically
nonzero for all temperatures. 
The average number of polymer-wall contacts at opposite
sides of an adsorption transition
behaves similarly to $\langle N_{AB}\rangle$  here.
In addition, like in adsorption, at $T=T_c$ we find that
$\langle N_{AB}\rangle / N = - \langle H\rangle /N \sim N^{\phi-1}$.
By fitting this behavior,
we obtained alternative $\phi$ estimates
($\phi = 0.55\pm 0.04$ and $\phi = 0.61 \pm 0.03$, in $2D$ and $3D$ respectively),
less sharp, but fully consistent with those based on the specific heat.

Apart from being a transition from swollen to compact state,
the diblock copolymer collapse has nothing in common with a homopolymer
theta collapse, which is characterized by a peculiar
$\nu$ exponent different from that of a high $T$ swollen chain
($\nu_{\Theta}=4/7$ and $1/2$, in 2D and 3D, respectively\cite{DS87,Carlo98}).
Also the theta point crossover exponent is
clearly different from the $\phi$'s here ($\phi_{\Theta}= 3/7 $ in 
$2D$\cite{DS87} and $\phi_{\Theta}=1/2$ in $3D$ \cite{Carlo98}).
The copolymer collapse is indeed a
novel transition, determined by the peculiar inhomogeneous
backbone structure and by its interactions. This and other remarkable
features discussed below, confer this transition a prototypical interest,
also beyond its possible applications. 
\begin{figure}
\vskip -0.6truecm
\inseps{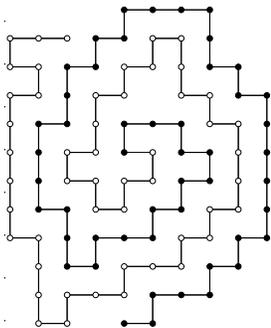}{0.22}
\caption{Typical configuration at low $T$ ($T=0.5 < T_c$) in d=2 for N = 100.
Filled and empty circles denote, respectively, A and B monomers.}
\label{fig4}
\end{figure}
Further insight
can be gained by considering the structure of the dense phase.
In both 2D and 3D, typical dense configurations generated
at low $T$
by our sampling technique
present a sort of maximal alternance of A and B
monomers, compatibly with the constraint of forming a diblock chain.
In 2D (Fig.~\ref{fig4}) A and B are also zipped on each other, and the 
resulting
double chain structure forms spirals, making the number of attractive 
interactions maximal. The zipped, double chain represents a concrete
realization of a polymer with orientation dependent interactions.
For models of such polymers, spiral compact phases have been recently observed
in 2D\cite{TS97}.

In 3D the compact phase has more complex configurations 
and substantial entropy. 
In this case zipping into a double chain does not occur.
On the other hand, the A and B monomers  are not segregated, 
and their alternance in space resembles, to some extent, 
an antiferromagnetic microphase structure.
By assuming $\nu_c=3/4$,
we can predict $\phi$ exactly in 2D, following
a strategy of identification with geometric percolative
objects, which in the past revealed extremely fruitful for homopolymer 
problems\cite{DS87,VSS91,SSV93}. The idea is that the statistics of suitably 
chosen percolative
contours could reproduce the critical behavior of interacting polymers in
the conditions of interest.
The homopolymer at the theta point in 2D is known to have
the same statistics as the external perimeter (hull) of a critical
percolation cluster\cite{CJMS87}, and this coincidence is at the 
basis of the full exact characterization of theta 
scaling\cite{DS87,VSS91,SSV93}.
On square lattice, the incipient infinite cluster of the percolation
problem for elementary squares ( only squares with an edge in common are
connected) is expected to have an externally
accessible hull which, besides assuming the configurations of a self-avoiding 
ring, has the statistical fractal dimension $D_{c1} =1/\nu_{SAW}=4/3$ of a swollen 
SAW\cite{GA86}. 

\begin{figure}
\inseps{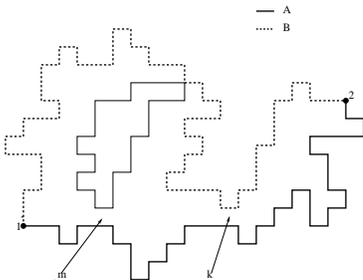}{0.25}
\caption{Points 1 and 2 separate the accessible hull portions A and
B. The light-continous line represents an inaccessible reentrance
of the hull. $k:$ red square corresponding to AB n.n. contacts.
$m:$ contacts of A with a hull reentrance.
The union of A and B has fractal dimension $4/3$. 
By including in the hull definiton also inaccessible reentrances, the
dimension becomes $7/4$.
} 
\label{fig6}
\end{figure}

This has been recently established on a  rigorous
basis in Ref.\onlinecite{ADA99}. By externally accessible
hull we intend that there should exist at least one path of connected
vacant squares by which any neighborhood of the
perimeter can be connected to points at infinite distance on
the lattice.
We try to identify our A and B blocks
at the transition with the two halves in which the externally 
accessible hull of the cluster
is divided by two diametrally opposite, distant points (Fig.~\ref{fig6}).
The identification of A and B with the copolymer blocks
at collapse will make sense
provided the reciprocal n.n. contacts of these halves have a positive
fractal dimension, matching the correct
$\phi$ exponent, as discussed below. This positive fractal
dimension is in fact the result of an effective attractive AB
interaction. 
As sketched in Fig.~\ref{fig6}, the fact that A and B are identified with
externally accessible hull portions excludes from the count the
full hull self-contacts possibly produced by the presence
of reentrances inaccessible from the exterior. Indeed, such
reentrances can not be included
in A or B, and it is in virtue of their exclusion
that the accessible perimeter has a fractal dimension $D_{c1}<D_{hull}=7/4$. 
%Contacts between the
%accessible hull portions A and B like those in
%Fig.~\ref{fig6} can be connected to infinity by at least two 
%distinct paths of vacant 
%squares. 
In Ref.~\onlinecite{ADA99} Coulomb gas results for the 
$T=0$ loop gas are used for a systematic determination of the dimensions of
various fractal sets in the cluster of
the hexagon percolation problem. For this problem
it turns out to be more easy to discuss geometry
and effective interactions of these sets.
In view of the peculiar connectivity 
properties of the
hexagonal lattice, hull reentrances are always
accessible by at least one path of vacant hexagons. 
Thus, external accessibility requires now
existence of at least two
distinct vacant paths joning the neighborhood of the hull to infinity~\cite{ADA99}.
The probability with respect to all percolative configurations
of a given hull profile, can be interpreted as
the result of a decision process (between being
occupied or vacant) for the hexagons which are progressively touched by
the hull itself. It is thus easy to realize that AB n.n. contacts
have the same effect as attractive interactions in a polymer
problem. Indeed, while normally each step of the hull involves
decision for a new hexagon (and thus a probability factor $1/2$,
the hexagon percolation threshold, in the total weight), at an AB 
contact the involved hexagon has already made his decision,
and no factor is required. Thus, AB contacts increase the
probability of the hull configuration.
Besides determining $D_{c1}$ exactly, the methods of Ref.~\onlinecite{ADA99}
allow us to conclude that the set of AB contacts described above
must have a fractal dimension $D_{c2}=3/4$, which coincides with
that of the ``red'' hexagons (or squares in our case) of the cluster,
i.e. of those hexagons whose removal would interrupt the cluster
connection between
the opposite distant points\cite{nota1}. $D_{c1}$ and $D_{c2}$ are
also expected to be universal with respect to different 
lattice structures\cite{ADA99}, and thus should hold also for square lattice.
The proposed identification leads 
to $\phi=9/16$: indeed,
for $N$-steps accessible half-perimeters, the average number
of contacts should scale like $N^{D_{c2}/D_{c1}}$. $D_{c2}/D_{c1}=9/16=0.56..$
is strikingly close to our numerical determination,
supporting the correctness of our assumptions\cite{nota}.

In summary, we gave a description of the collapse transition undergone
by a copolymer with two reciprocally attracting blocks. 
We could determine with high precision, and,
by an identification with percolation paths, exactly
in 2D, the universal exponents of the new transition, which
has analogies with homopolymer
adsorption and theta collapse,
but falls in a universality class different from both.
The low $T$ dense phase has antiferromagnetic features and,
in 2D, turns out to be spiral, with zipped A and B.
To our knowledge, the exact determination of $\phi$ in 2D is the first
application of percolation results to a genuinely
inhomogeneous polymer problem, and demonstates 
the potential of methods
which, about a decade ago, led to the correct characterization of
the homopolymer theta point.
We checked that the universality class of the transition  remains
the same if, more
consistently with the physical picture of oppositely charged blocks,
AA and BB repulsions are included in Eq.(1).
A natural extension of our study
consists of a systematic exploration of the phase diagram
in regions where, e.g., the AA and BB interactions  are also
attractive and compete with the AB one\cite{OSS99}. This
corresponds to a copolymer in which the difference between AA ( or BB)
and AB interactions is produced by hydrogen bonds.
We preliminarily verified that this competition
generates a very rich diagram, with both segregated and
unsegregated compact phases and interesting new transitions.

We thank Stu Whittington for useful discussions. 
A.L.S. acknowledges partial support from the European Network Contract
N. ERBFMRXCT980/83.

%%%%%% FIGURE

%\newpage
%\newpage
%\newpage
%\newpage
%\begin{figure}
%\vskip .2truecm
%\inseps{figure5.eps}{0.5}
%\vskip .6truecm
%\caption{Typical configuration at low $T$ in d=3 for N = 200. Light and
%heavy greys distinguish between A and B, respectively. }
%\label{fig5}
%\end{figure}
%\newpage

\end{multicols}


\begin{references}
\bibitem{nota1} Since also inaccessible hull reentrances can
produce red hexagons ( or squares, see Fig.5) we argue 
that these are, at most, a finite fraction of the total of 
red hexagons.
\bibitem{nota} Based on the identification
of a ring polymer at the theta point with the full cluster hull,
one can similarly establish that two halves of a homopolymer at the theta
point have a number of contacts $\sim N^{\phi_{\Theta}}$, where
$\phi_{\Theta}=3/7$.
\end{references}
\end{document}